\documentclass{eas}
\usepackage{graphicx}
\begin{document}

%%-----------------------------
%%      the top matter
%%-----------------------------

\title{Solar diameter with 2012 Venus transit} 

\runningtitle{Solar diameter with 2012 Venus transit}

\author{C. Sigismondi}
\address{Sapienza University of Rome and ICRA, University of Nice-Sophia Antipolis, IRSOL and GPA-Observatorio Nacional Rio de Janeiro. e-mail: sigismondi@icra.it}

\begin{abstract}

The role of Venus and Mercury transits is crucial to know the past history of the solar diameter. Through the W parameter, the logarithmic derivative of the radius with respect to the luminosity, the past values of the solar luminosity can be recovered.
The black drop phenomenon affects the evaluation of the instants of internal and external contacts between the planetary disk and the solar limb. With these observed instants compared with the ephemerides the value of the solar diameter is recovered. 
The black drop and seeing effects are overcome with two fitting circles, to Venus and to the Sun, drawn in the undistorted part of the image.
The corrections of ephemerides due to the atmospheric refraction will also be taken into account.  
The forthcoming transit of Venus will allow an accuracy on the diameter of the Sun better than 0.01 arcsec, with good images of the ingress and of the egress taken each second. 
Chinese solar observatories are in the optimal conditions to obtain valuable data for the measurement of the solar diameter with the Venus transit of 5/6 June 2012 with an unprecedented accuracy, and with absolute calibration given by the ephemerides.
 
\end{abstract}

\maketitle

%%-----------------------------
%%      your text
%%-----------------------------

\section{The method of eclipses}

I. I. Shapiro in 1980[\cite{shapiro}] used records of transits of Mercury to recover the past history of the solar diameter.[\cite{svesh}]
Further studies seems to confirm the constancy of the diameter within the errorbars.
Measurements made with different instruments, under perfect observing conditions, as in the case of Gambart and Bessel in 1832 yield different transit times, and different diameter of Mercury and, consequently, a different diameter of the Sun.[\cite{Gambart,Bessel}]

The determination of the planetary diameter is subjected to the Point Spread Function of the telescope matching with the Limb Darkening Function of the Sun,[\cite{golub}] and, in the case of Venus, there is also the atmosphere. 
 
Nowadys the chord draft on the solar limb by the planet's disk can be recovered by photos, in conditions not affected by black drop phenomenon.[\cite{sigi,UAI}]
The time in which the chord is zero, when the black drop is maximum, can be extrapolated from UTC labelled photographs made each second around the intermediate stages of ingress and egress.
After corrections for Earth's atmospheric refraction the ephemerides can be used to recover the solar diameter by comparison with the observed ingress and egress times.
The opportunity given by the forthcoming transit of Venus of 5/6 june 2012 and the one of Mercury of May 9, 2016 has to be exploited to measure the solar diameter with unprecedented accuracy.
The studies on the Venus aureole[\cite{tanga}] if done with UTC synchronized high-resolution photos, can be useful to do solar diameter measurements, once the location of the observations are known with GPS coordinates.

\begin{figure}
\centerline{\includegraphics[width=1\textwidth,clip=]{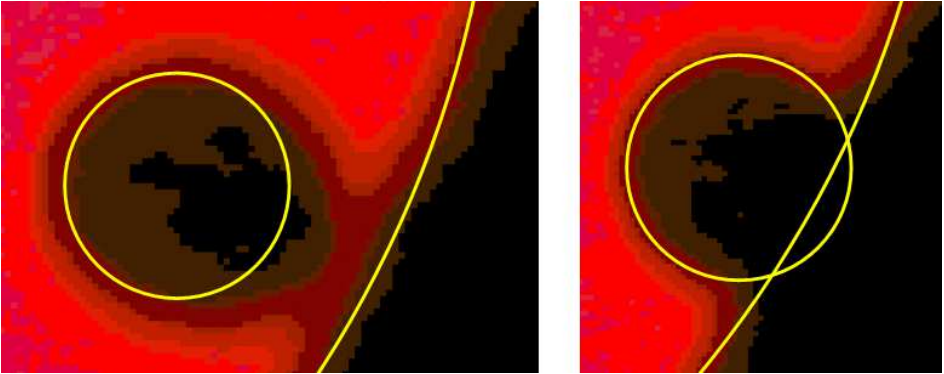}}
\caption{Two stages of the egress of Venus in 2004 (H$\alpha$ images of Anthony Ayomamitis, Athens, details). The circular profiles of Venus and Sun are fitted to the undistorted part of the image. }
\label{Fig. 1}
\end{figure}

\section{Conclusions: one image per second during 2012 ingress/egress}
We study the position of the center of the planet with respect to the inflexion point of the limb darkening function of the Sun (the solar lim). By using the circular fits to the undistorted part of the image we avoid
the black drop effect. We can use two reference instants of comparison with ephemerides: 
\begin{itemize}
\item{the time when the center of Venus crosses the solar limb.}
\item{the time when the chord drawn by Venus and the solar limb becomes zero.[\cite{UAI}]}
\end{itemize}
The first method can be applied also to the external aureole[\cite{tanga}] produced by the refraction in the upper atmosphere of Venus, and this method is independent on the thickness of the atmosphere,  because it uses the center of the planet. Similarly the determination of the center of the planet is less affected by the black drop phenomenon with respect to the second method of the length of the chord. 

The method of the chord has been tested with 50 images made each minute, 25 at ingress and 25 at egress, in the H$\alpha$ line by A. Ayomamitis with a 16 cm apochromatic refractor near Athens during the Venus transit of 2004: the accuracy without refraction correction has been 2.6 s at the egress and 8.1 s at the ingress (with the Sun low near the horizon); the final accuracy on the whole solar radius was 0.38 arcsec.

An improvement on the final accuracy is expected with a 1 s photo sampling (60 times larger; accuracy $0.38/\sqrt{60}\sim 0.05$ arcsec), and a further improvement will come from the atmospheric refraction corrections and from the determination of the center of the planet, instead of the chord's length.
The ultimate level of accuracy below 0.01 arcsec is the goal of this measurement: the most precise achievable with ground based methods.
China is under optimal conditions to observed the transit of Venus of 2012, and to gather useful images to accurately measure the actual solar diameter.

\thanks{\bf Acknowledgments}
Thanks to Anthony Ayomamitis for the images of the Venus transit of 2004. Despite of the worldwide campaign to observe this lifetime phenomenon nobody else (amateurs and observatories) yet published a sequence of  chronodated images useful to measure the solar diameter by using the transit of Venus. 
Thanks to Patrick Rocher (IMCCE) for a fruitful discussion on the ephemerides.
%
%-----------------------------
%%      your bibliography
%%-----------------------------

\end{document}